\documentclass[12pt,preprint]{aastex}

\newcommand{\HI}{H~{\sc i}} 

\newcommand{\kms}{${\rm km~s^{-1}}$}

\shortauthors{MCCLURE-GRIFFITHS ET AL} 
\shorttitle{GALACTIC CHIMNEY: GSH 242-03+37}

\begin{document} 

\title{Evidence for chimney breakout in the Galactic supershell GSH 242-03+37}

\author{N.\ M.\ McClure-Griffiths,\altaffilmark{1} Alyson
  Ford,\altaffilmark{1,2} D.\ J.\ Pisano,\altaffilmark{3,4} B.\ K.\
  Gibson,\altaffilmark{2,5,6} L.\ Staveley-Smith,\altaffilmark{1}
  M.\ R.\ Calabretta,\altaffilmark{1} L.\ Dedes,\altaffilmark{7} and
  P.\ M.\ W.\ Kalberla\altaffilmark{7}}

\altaffiltext{1}{Australia Telescope National Facility, CSIRO, PO Box 76, Epping
 NSW 1710, Australia; naomi.mcclure-griffiths@csiro.au} 
\altaffiltext{2}{Centre for Astrophysics and Supercomputing, Swinburne University of Technology, Hawthorn VIC 3122, Australia}
\altaffiltext{3}{Remote Sensing Division, Naval Research Laboratory,
  Code 7213, 4555 Overlook Ave. SW,
 Washington, D.C, 20375 }
\altaffiltext{4}{National Research Council Postdoctoral Fellow}
\altaffiltext{5}{Laboratoire d'Astrophysique, Ecole Polytechnique Federale de Lausanne
(EPFL), Observatoire, CH-1290, Sauverny, Switzerland }
\altaffiltext{6}{Centre for Astrophysics, University of Central Lancashire, Preston,
PR1 2HE, United Kingdom}
\altaffiltext{7}{Radioastronomisches Institut der Universit\"at Bonn,
 Auf dem H\"ugel 71, 53121 Bonn, Germany}

\authoraddr{Address correspondence regarding this manuscript to: 
                N. M. McClure-Griffiths
                ATNF, CSIRO
                PO Box 76
		Epping NSW 1710
		Australia }
\begin{abstract}
We present new high resolution neutral hydrogen (\HI) images of the
Galactic supershell GSH 242-03+37.  These data were obtained with the
Parkes Radiotelescope as part of the Galactic All-Sky Survey (GASS).
GSH 242-03+37 is one of the largest and most energetic
\HI\ supershells in the Galaxy with a radius of $565\pm 65$ pc and an
expansion energy of $3 \times 10^{53}$ ergs.  Our images reveal a
complicated shell with multiple chimney structures on both sides of
the Galactic plane.  These chimneys appear capped by narrow filaments
about 1.6 kpc above and below the Galactic mid-plane, confirming
structures predicted in simulations of expanding supershells.  The
structure of GSH 242-03+37 is extremely similar to the only other
Galactic supershell known to have blown out of both sides of the
plane, GSH 277+00+36.  We compare the GASS \HI\ data with X-ray and
H-$\alpha$ images, finding no strong correlations.

\end{abstract}
\section{Introduction}
\label{sec:intro}
Among the most dramatic structures in the interstellar medium (ISM) of
disk galaxies are large shells and supershells.  These objects are
generally observed as voids in the neutral hydrogen (\HI) distribution
surrounded by swept-up walls. In some nearby galaxies, like the Large
and Small Magellanic Clouds, the \HI\ structure of the disk is
dominated by shells and supershells
\citep{kim99,hatzidimitriou05}. The ISM of the Milky Way is also
riddled with tens, if not hundreds, of \HI\ shells
\citep[e.g.][]{heiles79,heiles84,mcgriff02a,ehlerova05}. It is thought
that most \HI\ shells are formed by stellar winds or supernovae, or
the combined effects of both. This explanation is particularly
convincing for smaller shells, with diameters of a few tens of parsecs
and formation energies on the order of $10^{52}$ ergs. However, the
larger shells, or supershells, are enigmatic.  They seem to require
unreasonably large ($>10^{53}$ ergs) formation energies in order to
maintain expansion velocities of $\sim 20$ \kms\ at radii in excess of
a few hundred parsecs \citep{heiles84}.  The stellar wind and
supernova from a given massive star is capable of injecting $\sim
10^{51}$ ergs of energy into the ISM, suggesting that many hundreds
and even thousands of massive stars are required to power the most
energetic shells.  In this case, it is expected that multiple
generations of star formation are required, and although numerous
studies have searched for evidence of triggered star formation
associated with \HI\ shells there are few examples \citep{oey05}.

\HI\ supershells play a significant role in the energy budget of the
ISM and can also play a role in the exchange of matter between the
disk and halo. \HI\ supershells can grow large enough to exceed the
scale height of the \HI\ disk.  In this case, the shell expands
rapidly along the density gradient away from the disk until it becomes
unstable and breaks out, venting its hot internal gas to the halo.
This ``chimney'' process may provide a mechanism for distributing hot
gas and metals away from the disk \citep*{dove00}.  \HI\ shell
blow-outs are predicted for, and indeed observed in, a number of  dwarf galaxies
where the gravitational potential of the disk is smaller than in large
spirals \citep{maclow99,marlowe95}.  Occasionally chimneys are 
observed in large spiral galaxies, with good examples being
NGC 891, NGC 253 and NGC 6946 \citep{rand90,howk97,boomsma05}.  In the
Milky Way, very little is known about the impact of chimneys on the
formation, structure and dynamics of the halo.  In fact, only a
handful of relatively small chimneys are known in the Milky Way,
e.g. the Stockert chimney \citep{muller87}, the W4 chimney
\citep{normandeau96,reynolds01}, the Scutum supershell
\citep{callaway00} and GSH 277+00+36 \citep{mcgriff00}. Together, the
known chimneys are not capable of providing the thermal energy
required to support the halo.

One of the largest \HI\ supershells in the Milky Way is GSH 242-03+37,
discovered by \citet{heiles79} in his seminal work on Galactic shells.
GSH 242-03+37 is located at $l=242\arcdeg$, $b=-3\arcdeg$, which is in
the direction of the so-called ``Puppis window'', an area of very low
visual extinction \citep{fitzgerald68}. The shell has an angular
diameter of $15\arcdeg$.  Its kinematic distance is 3.6 kpc, implying
a physical diameter of $\sim 1$ kpc.  \citet{heiles79} suggested that
GSH 242-03+37 is still expanding with an expansion velocity of $v_{\rm
  exp} \approx 20$ \kms\ and from that he estimated an expansion
energy of $E_E \sim 1.6 \times 10^{54}$ ergs.  Despite the impressive
size and implied energetics of this shell, there has been very little
follow-up work.  \citet{stacy82} made a thorough study of the \HI\ in
the region $239\arcdeg \leq l \leq 251\arcdeg$ with the aim of
correlating \HI\ features with optical spiral tracers.  The survey
concentrated mainly on smaller scale \HI\ features, and while it
mentioned GSH 242-03+37, no detailed images were available or
discussed.

Here we use new data from the Galactic All-Sky Survey (GASS) to study
the \HI\ supershell GSH 242-03+37.  We present the highest angular
resolution ($\sim 15\arcmin$) and most sensitive images of the shell
published so far.  We show that GSH 242-03+37 has in fact broken out of
both sides of the Galactic plane through three large channels.  We
show that these chimney openings are capped at high Galactic latitude
with very narrow, low surface brightness filaments.  We discuss the
chimney caps and their long-term fate in \S \ref{sec:chimneys}.  In
\S\ref{subsec:otherwavelengths} we examine archival X-ray (ROSAT) and
H-alpha (SHASSA) data to explore the various gas phases associated
with the chimney.  In \S \ref{sec:stars} we discuss the stellar
content of the shell and in \S \ref{sec:minivoid} we explore the
possible association of a small shell that appears inside GSH
242-03+37.  

\section{Observations and Analysis}
\label{sec:obs} 
The \HI\ data presented here are from the first pass of the Parkes
Galactic All-Sky Survey (GASS; McClure-Griffiths et al., 2005, in
prep.).  GASS is a project to image \HI\ at Galactic velocities
($-400~{\rm km~s^{-1}} \leq v_{\rm LSR} \leq +450~{\rm km~s^{-1}}$) for
the entire sky south of declination $0\arcdeg$.  GASS uses the
Parkes multibeam to produce a fully sampled atlas of \HI\ with an
angular resolution of $15\arcmin$, spectral resolution of $0.8~{\rm
  km~s^{-1}}$, and to an rms sensitivity of 80-90 mK.  The survey will be
corrected for stray radiation effects, according to the method
described in \citet{kalberla05}, to ensure high reliability of the
\HI\ spectra.  Observations for the survey began in January 2005 and
will continue through to 2007.  When complete, GASS will be the first
fully sampled all-sky survey of \HI\ on sub-degree scales.  The full
survey details, including its scientific goals, will be described in a
future paper.  Here we briefly describe the observations and data
reduction techniques to allow assessment of the data presented.

GASS is conducted as an on-the-fly mapping survey, with each point in
the sky scanned twice.  We use the Parkes multibeam
\citep{staveley-smith96}, which is a thirteen beam receiver package
mounted at prime focus on the Parkes Radiotelescope near Parkes NSW,
Australia.  The thirteen beams of the multibeam are packed in a
hexagonal configuration with a beam separation on the sky of
$29\farcm1$ for the inner beams.  On-the-fly mapping is performed by
scanning the telescope at a rate of 1~deg~min$^{-1}$, recording
spectra every 5 seconds.  While scanning, the receiver package is
rotated by $19\fdg1$ with respect to the scan direction to ensure that
the inner seven independent beams make parallel tracks equally spaced
by $9\farcm5$ on the sky.  Scans in both right ascension and
declination will be made for the full survey, although only a few RA
scans have been included in this paper.  The declination scans are made at a
constant RA and are 8 deg long in declination.  After a scan the
receiver package is offset in RA to perform an interleaved scan,
reducing the spacing between adjacent beam tracks to $4\farcm7$. 

Spectra are recorded in a special correlator mode that allows 2048
channels across an 8 MHz bandwidth on all thirteen beams.  In-band
frequency switching is used to allow for robust bandpass correction.
We switch every 5 seconds between center frequencies of 1418.8345 MHz
and 1421.9685 MHz.  Bandpass calibration is done in near real-time
using the {\em Livedata} package, which is part of the ATNF subset of
the {\em aips++} distribution.  The bandpass correction algorithm
employed was designed expressly for the GASS frequency-switched data.
It works on each beam, polarization and IF independently, performing a
robust polynomial fit to the quotient spectrum (one frequency divided
by the second frequency) after masking the emission by examining the
spectrum both spectrally and spatially.  {\em Livedata} also performs the
Doppler correction to shift the spectra to the Local Standard of Rest
(LSR).  Absolute brightness temperature calibration was performed from daily
observations of the IAU standard line calibration regions S8 and S9
\citep{williams73}.

Calibrated spectra are gridded into datacubes using the {\em
  Gridzilla} package, also part of the ATNF subset of the {\em aips++}
distribution.  The gridding algorithm used in {\em Gridzilla} is
described in detail in Barnes et al.\ (2001)\nocite{barnes01}.  GASS
spectra were imaged using a weighted median technique with a cell size
of $4\arcmin$, a Gaussian smoothing kernel with a full width half max
of 12\arcmin, and a cutoff radius of $8\arcmin$.  The effective
resolution of the gridded data is $\sim 15\arcmin$.  The per channel
rms of the resulting image cubes near the Galactic plane is $\sim 120$
mK.  These data are not corrected for stray radiation and may
therefore contain some low-level spurious features.  For the data
presented here we have compared our images with the low resolution
stray radiation corrected Leiden/Argentine/Bonn survey
\citep{kalberla05,bajaja05} to verify features.

\section{Results}
\label{sec:results} 
\citet{heiles79} cataloged GSH 242-03+37 with a low confidence
rating, suggesting uncertainty about the shell's veracity.  With the
improved angular and spectral resolution of the GASS data, as well as
the availability of improved data visualization tools, we are
confident that this shell meets the three criteria for shell
identification given in \citet{mcgriff02a}, i.e.\ that the void is
well-defined over more than three consecutive velocity channels with
an interior to exterior brightness contrast of 5 or more, that the
void changes shape with  velocity and that a velocity profile
through the shell shows a well-defined dip flanked by peaks. 

\begin{figure}
\centering
\includegraphics[width=4.5in]{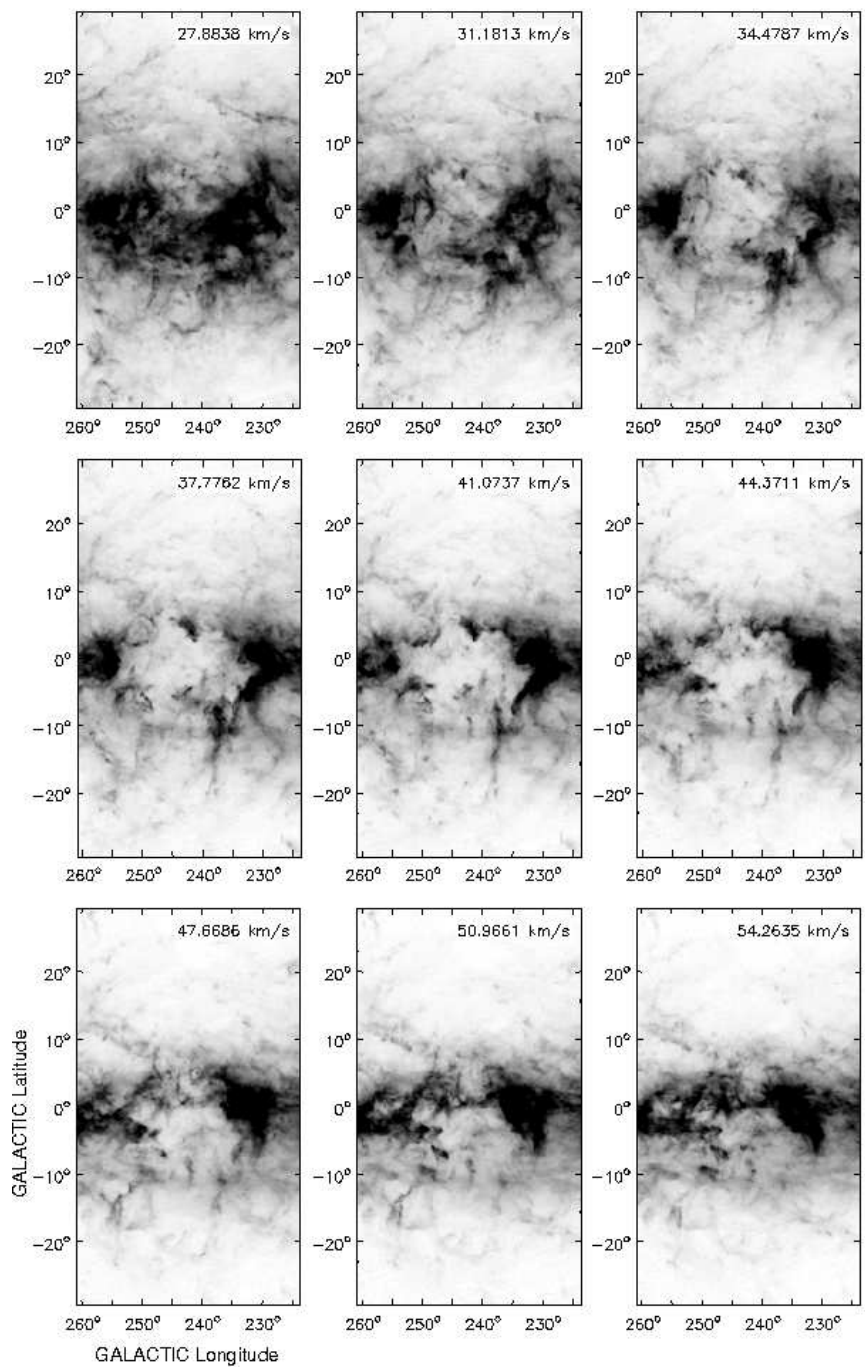}
\caption[]{\HI\ velocity channel images of GSH 242-03+37.  Every
  fourth velocity channel between $v=27.9$ \kms\ and $v=54.3$ \kms\ is
  displayed.  The LSR velocity  of each channel is shown in
  the upper right corner of each panel.  The grey scale has a scaling
  power cycle of $-1$ and goes between 0 (white) and 50 K (black).  The
  expanding shell develops slowly from the first panel to its widest
  point near $v=42$ \kms.  The rear wall is apparent near $v=54$ \kms.
\label{fig:hishell}}
\end{figure}

In Figure \ref{fig:hishell} we show velocity channel images of the
shell as multiple panels.  The shell is visible as the large void in
the center of the images, between LSR\footnote{All velocities are
  quoted with respect to the kinematic Local Standard of Rest.}  velocities
$v\approx 30$ \kms\ and $v\approx50$ \kms.  Every fourth velocity
channel is displayed here to give an impression of the dynamic
structure in the shell.  The first and last panels of
Fig.~\ref{fig:hishell} show the approximate front and back caps of the
shell.  The shell extends over approximately 18 degrees in longitude
and 10 degrees in latitude.  However, there are clear breaks on the
top and bottom of the shell, as seen in the velocity channel images.
These are indicative of chimney openings and will be discussed
thoroughly below.  We find that the center of the shell is at a
slightly different location than given in \citet{heiles79}.  We define
the center as the velocity of least emission in the spectral profile
through the shell center and the geometric center of the shell at that
velocity.  Using these criteria, the center of the shell is at
$l=243\arcdeg$, $b=-1.6\arcdeg$, $v=+42$ \kms, notably different than
the coordinates implied by its name.

\begin{figure}
\centering
\plotone{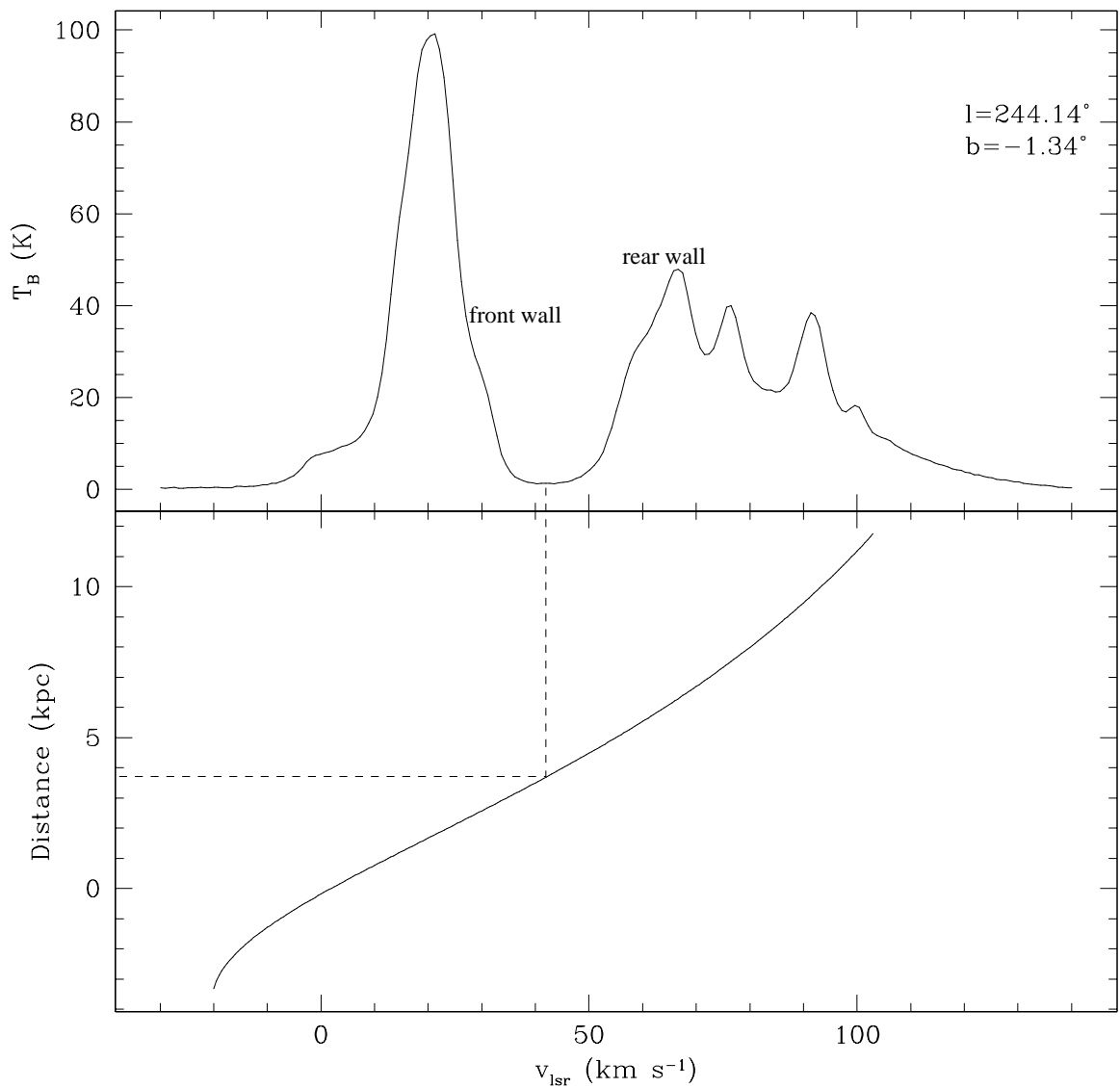}
\caption[]{Velocity profile through the center of GSH 242-03+37,
  showing the deep well of the shell at $v=42$ \kms\ and the peaks for
  its front and rear caps.  The profile was extracted at
  $l=244\fdg14$, $b=-1\fdg34$.  The bottom panel shows the
  velocity-distance relation for this line of sight as defined by the
  \citep{brand93} rotation curve.  The kinematic distance of the shell
  center is at $d \approx 3.6$ kpc.
\label{fig:profile}}
\end{figure}

A velocity profile through the shell center is shown in the top panel
of Figure \ref{fig:profile}.  The shell is the clearly defined dip in
the profile at $v\approx 40$ \kms.  The front and back caps of the
shell are marked on Fig.\ \ref{fig:profile} and are apparent as the
bumps in the velocity profile at $v\approx 27$ \kms\ and $v\approx 57$
\kms\ on both sides of the void.  The shell is located at a Galactic
longitude where the rotation curve is relatively simple, allowing us
to translate radial velocity approximately into distance.  The lower
panel of Fig.\ \ref{fig:profile} plots the velocity-distance relation
at $l=244\arcdeg$ from the \citet*{brand93} rotation curve, assuming
the IAU recommended values for the Galactic center distance, $R_0=8.5$
kpc, and LSR velocity, $\Theta_0 = 220$ \kms.  From this relationship
the kinematic distance of the shell is 3.6 kpc, as was also found by
\citep{heiles79}.  The shell is at a Galactocentric radius of $R_{\rm
  g} = 10.7$ kpc.  The error on the kinematic distance is on the order
of 10\% because of uncertainties in determining the central velocity
of the shell, random cloud-to-cloud motions in the ISM and errors in
the rotation curve.  The radius of the shell along the plane is
$R_{\rm sh} = 565 \pm 65$ pc. At a distance of 3.6 kpc our resolution
is approximately 16 pc.  Because of the large size of the shell it may
be elongated because of differential rotation in the Galactic plane
\citep{tenorio88}.  This will distort the shell and affect its
lifetime as discussed below.

The interior of the shell has extremely low brightness temperature
values when compared with the rest of the Galactic plane; the mean
brightness temperature in the shell interior is only $T_{\rm b} = 4$ K
with a standard deviation on the mean of $\sigma_{\rm T} = 1.5$ K.
Assuming the gas is optically thin, this implies a mean column density
of $N_{\rm H} = 1.3 \pm 0.6 \times 10^{20}~{\rm cm^{-2}}$ and a mean
internal \HI\ number density of $n_{\rm H}\sim 0.07~{\rm cm^{-3}}$ if
the shell is spherical.

\subsection{GSH 242-03+37 Physical Properties}
The physical properties of GSH 242-03+37, such as radius, mass,
expansion velocity, and expansion energy were estimated by
\citet{heiles79}.  Our values are only slightly different from those.
Both our values and Heiles' values, where different, are given in
Table \ref{tab:params}.

Expansion velocities for shells are usually estimated as half of the
total measured velocity width, $\Delta v$, of the shell. The full
velocity width of GSH 242-03+37, through the center of the shell, is
approximately $\Delta v=25$ \kms.  Because of the relationship between
distance and radial velocity, there is a complicated coupling of the
expansion velocity, $v_{\rm exp}$, and the velocity width due to the
line-of-sight physical dimension of the shell, $v_p$.  A simplistic
way of de-coupling the expansion velocity and velocity width due to
physical size is to use the velocity gradient, $dv/dr$, to estimate
the contribution of the physical size to the total velocity width.
Again using the \citet*{brand93} rotation curve, we find that at
$v=42$ \kms\ along this line of sight $dv/dr \sim 10~{\rm
  km~s^{-1}~kpc^{-1}}$.  If the diameter of the shell along the
line-of-sight is comparable to its diameter in the plane of the sky,
then $v_p \sim 10~{\rm km~s^{-1}}$. We then make the simplifying
assumption that the total velocity width is $\Delta v \approx 2v_{exp}
+ v_p = 25$ \kms, implying $v_{\rm exp} \approx 7$ \kms.  

The expansion energy, $E_E$, of a shell is defined by \citet{heiles79}
to be the equivalent energy that would have been deposited at the
center of the shell to account for the observed radius and expansion.
The expansion energy, based on the calculations of \citet{chevalier74}
for supernova expansion, is $E_E = 5.3 \times
10^{43}\,n_0^{1.12}\,R_{\rm sh}^{3.12}\,v_{\rm exp}^{1.4}$, where
$n_0$ is the ambient density measured in ${\rm cm^{-3}}$, $R_{\rm sh}$
is in parsecs, $v_{\rm exp}$ is in \kms.  This equation makes the
extreme simplifying assumption that the ambient medium, $n_0$, into
which the shell is expanding, is homogeneous with constant density. We
know that this cannot be true on small scales, but for very large
shells the density variations largely average out and the equation
provides a reasonable standard energy estimate with which to compare
shells.  For GSH 242-03+37, assuming $n_0 \approx 1~{\rm cm^{-3}}$,
the expansion energy is $E_E\sim 3.1\times 10^{53}$ ergs.  Another
limitation of the expansion energy equation is that it does not
account for energy lost to high latitudes by shell break-out.
Therefore, for GSH 242-03+37 the expansion energy is only a lower
limit.  We note that our expansion energy estimate is about a factor
of 10 lower than the value quoted in \citet{heiles79}.  The difference
between our value and Heiles' can be accounted for by our assumption
of an ambient density of $n_0 \approx 1~{\rm cm^{-3}}$, whereas
\citet{heiles79} uses $n_0 \approx 2~{\rm cm^{-3}}$, and from the
lower expansion velocity estimated here.  If the average energy output
of a single O or B star via its stellar winds and supernovae is $\sim
10^{51}$ ergs, then more than 300 massive stars are required to expand
the shell to its current size.  There are no known coveal stellar
clusters of that size in the Milky Way, which suggests that GSH
242-03+37 was formed through the effects multiple generations of
massive stars.

Age estimates for \HI\ shells are also fraught with large
uncertainties.  Unless a powering source, such as an OB association,
that can be aged independently is associated with the shell it is
often impossible to accurately estimate the age of a shell.  We can,
however, estimate a shell's dynamic age based on models of the
evolution of supernova remnants in the late radiative phase.  In this
case the dynamic age, $t_6$ in units of Myr for a shell of radius,
$R_{\rm sh}$, given in pc and $v_{\rm exp}$ given in units of \kms, is
given by $t_6 = 0.29 \,R_{sh}/v_{\rm exp}$ \citep{cioffi88}.  For GSH
242-03+37, the dynamic age is $t\sim 21$ Myr.  Comparing with other
known Galactic shells, GSH 242-03+37 is relatively old
\citep{heiles79,heiles84,mcgriff02a}.  Ultimately the lifetime of
\HI\ shells is limited by the development of instabilities along
their walls and the onset of deformation and shear due to differential
rotation in the Galactic plane.  Both effects become significant at
around 20 Myr \citep{dove00,tenorio88}.  Differential rotation will
distort the shell so that it no longer appears spherical.  At a
Galactocentric radius of 10.3 kpc this is a moderate effect; over 20
Myr a static shell of radius 1 kpc will distort to have an axial ratio
in the plane of $\sim $1.5:1.

\subsection{GSH 242-03+37 Morphology}
\label{sec:morphology}
The three-dimensional morphology of GSH 242-03+37 is extremely
interesting. One of the noteworthy aspects of the morphology is that
the shell is not spherical, but has some ``scalloped'' structure along
the edges as can be seen in Figures \ref{fig:hishell} and
\ref{fig:annotate}.  The bases of these scallops are separated by
several degrees or arcmin, or $\sim 400-500$ pc.  In three locations,
two at the bottom and one at the top, these arches are weaker or
absent presenting extensions from the Galactic plane towards the halo.
The three break-outs are at: $(l,b)=(245\fdg2,+3\fdg8)$, $(243\fdg0,
-8\fdg3)$ and $(236\fdg5,-8\fdg2)$.  These break-outs or ``chimneys''
have some vertical structures that clearly separate them from the
ambient medium.  At very low brightness temperatures ($\sim 1.5$ K)
the chimney structures are capped, each about 1.6 kpc from the center
of the shell.  These caps are marked on Figure \ref{fig:annotate},
which is a channel image at $v=+45$ \kms.  They are also visible in
the channel images shown in Figure \ref{fig:hishell}.  Like the shell,
the caps are visible over $\sim 20$ \kms\ of velocity space, which
suggests that they are not only associated with the shell, but also
physically extended.  The structure and nature of these caps will be
discussed in \S \ref{sec:chimneys}.

\begin{figure}
\centering
\plotone{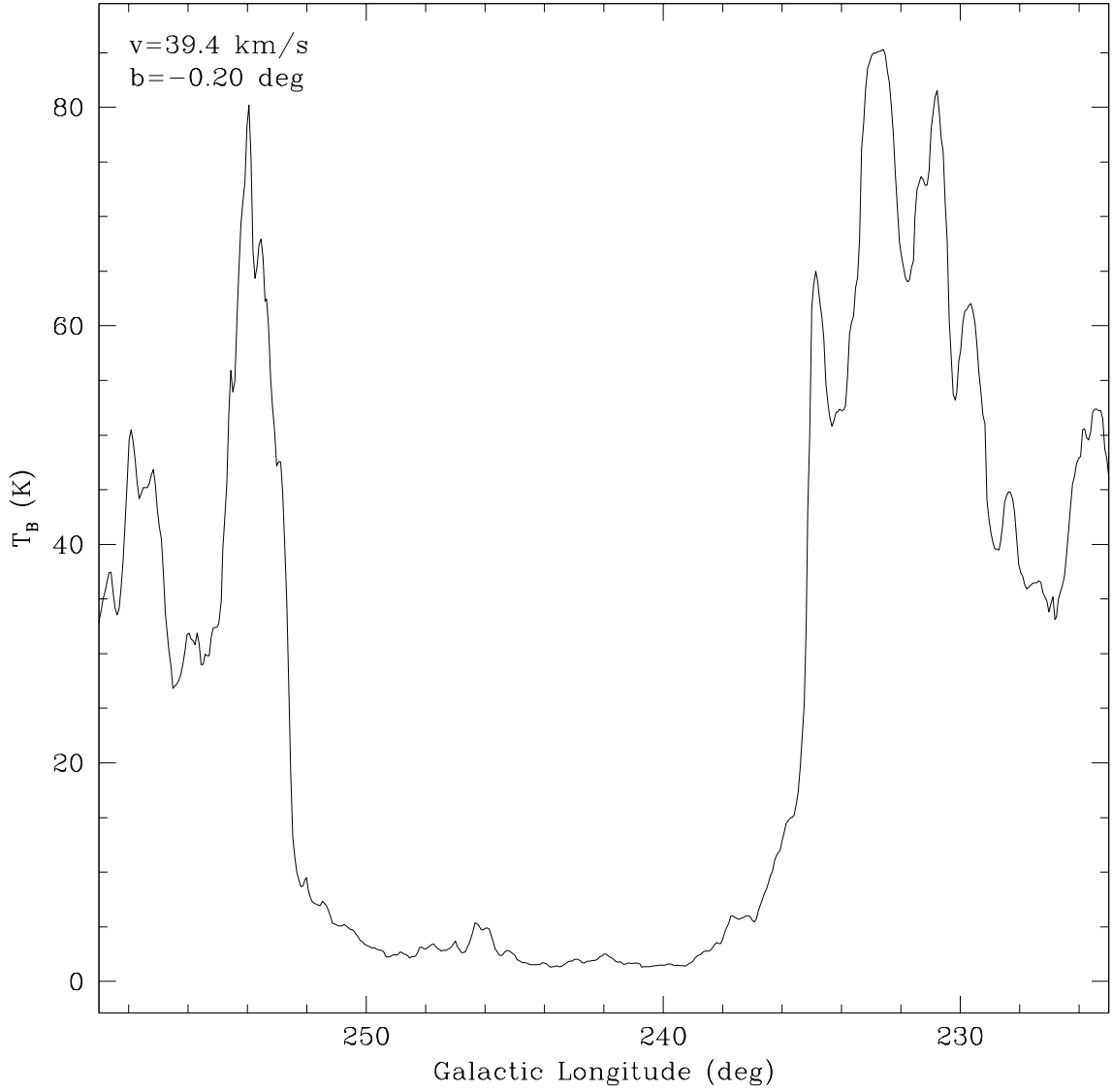}
\caption[]{Galactic longitude slice across GSH 242-03+37 at
  $b=0\fdg20$, $v=39.4$ \kms\ showing the sharp walls and deep void of
  the shell center.  The interior of the shell is quite empty with
  mean brightness temperatures of $T_b = 4\pm 1.5$ K with low level
  structure.  The sharp walls are indicative of a shell formed by compression.  
\label{fig:walls}}
\end{figure}

Figure \ref{fig:walls} shows a slice across the shell in the
longitudinal direction.  The slice is taken at $v=39.4$ \kms,
$b=0\fdg20$.  The shell is clearly  empty and the walls
of the shell are very sharp.  The walls show a brightness temperature
contrast of 10 to 20 from the shell interior to the shell wall over
one to two resolution elements, or $\sim 16 - 32$ pc.  Referring to
similarly strong shell walls in GSH 277+00+36, \citet{mcgriff03b}
suggested that the sharpness of the walls is indicative of compression
as associated with a shock. 

\citet{stacy82} noted one cloud in the shell interior at $(l,b,v)$ =
$(242\fdg2, -4\fdg6,36~{\rm km~s^{-1}})$, pointing out that it was
unusual as one of the only clouds within a largely evacuated area.
With the sensitivity of GASS it is clear that there is quite a lot of
structure inside the shell, though in general it is only at the $T_b
\sim 5$ K level.  There are a number of interlocking rings throughout
the shell interior.  Most of these are relatively circular and
noticeable over several velocity channels.  The cloud cataloged by
\citet{stacy82} appears to belong to a thin ring structure near the
center of GSH 242-03+37, as shown in Figure \ref{fig:minishell}.  This
ring is distinguished from the rest of the internal structure because
it encloses an interesting region that is even more evacuated than the
rest of the shell, appearing as a shell within the shell. This
``mini-shell'' is centered at $(l,b,v)$ = $(242\fdg9,-2\fdg3,45~{\rm
  km~s^{-1}})$ \kms\ and has an angular diameter of $\sim 3\fdg8$.
The interior has lower brightness temperatures than the main void; the
brightness temperatures associated with mini-shell are $\sim 2$ K
inside the void, about a factor of two lower than in the main void,
and $\sim 7 - 10$ K along the ``walls''.  There are few regions in the
Galactic plane that show such low brightness temperatures and most are
associated with known \HI\ shells.

Another particularly noticeable feature is the compact cloud a
$(l,b,v)$ = $(240\fdg9,+4\fdg9,45~{\rm km~s^{-1}})$, which can be seen in
several of the channel images presented in Figure~\ref{fig:minishell}.
The unresolved cloud is surrounded by a ring of emission with a
diameter of 2\arcdeg.  Also centered on this position, about 3 degrees
away, is an arc of emission.  These features, though noticeable, have
no obvious physical explanation.

\subsection{Comparison with other wavelengths}
\label{subsec:otherwavelengths}
We have obtained publicly available data from the ROSAT all-sky survey
maps of the diffuse X-ray background \citep{snowden97}.  We compared
the $1/4$ keV, $3/4$ keV and 1.5 keV emission with the
\HI\ distribution.  There is clear evidence for $\frac{1}{4}$ keV
excess emission in the shell interior.  It would be surprising,
however, if this excess were physically associated with shell.  At
$1/4$ keV, unity optical depth corresponds to \HI\ column densities of
about $1\times 10^{20}~{\rm cm^{-2}}$ \citep{snowden97}, giving a mean
free path for $1/4$ keV X-rays of only $\sim 65$ pc.  Even though the
\HI\ column density through the Puppis window is low, at the distance
of GSH 242-03+37 the foreground \HI\ column density is significant at
$\sim 2 \times 10^{21}~{\rm cm^{-2}}$.  Examining the \HI\ data cube,
we find that the morphology of the X-ray excess agrees very well with
the foreground \HI\ shell, GSH 242-04-05 \citep{heiles79}.  It is
therefore likely that the $1/4$ keV X-ray excess emission traces hot
gas in the foreground object, not in GSH 242-03+37.

The mean-free path for X-rays in the higher energy bands is much
longer; 1.5 keV electrons have a mean free path of $\sim 3$ kpc
\citep{snowden97}. Despite the potential detectability of higher
energy X-rays, we find no correlation between the \HI\ distribution
and the $3/4$ or 1.5 keV emission. The lack of harder X-ray
features is not unexpected, however, because $3/4$ and 1.5 keV
emission suggests very hot gas with temperatures on the order of
$10^7$ K.  Old large shells, like GSH 242-03+37 are not expected to
have gas temperatures much higher than about $3.5 \times 10^6$ K
\citep{maclow88}.

We have also compared the \HI\ distribution to the velocity integrated
H$\alpha$ emission from the SHASSA survey of the Southern sky
\citep{gaustad01}.  Because of the low extinction towards this shell
it should be possible to detect H$\alpha$ emission to the 3.6 kpc
distance of the shell.  Unfortunately, because there is no velocity
discrimination in the SHASSA data it is very difficult to distinguish
foreground H$\alpha$ features from features at the distance of the
shell.  We find very few obvious correlations between the \HI\ at
$v=35-50$ \kms\ and H$\alpha$ over the entire field of the shell.  The
only potential candidate for agreement is a faint H$\alpha$ filament
that lies just inside the \HI\ mini-shell at $l=243\arcdeg$, $b=-2\fdg3$.
This feature, however, is in a confused region and no definitive
association can be made.

Finally, we have searched the FUSE catalog of O{\sc vi} detections
(Wakker at al. 2003) to determine if there is hot, high-latitude gas
at the same velocity as GSH 242-03+37. Two O{\sc vi} detections below
the plane of the Galaxy are near GSH 242-03+37: the first towards PKS
0558-504 at 11 \kms\ and the second towards NGC 1705 at 29
\kms. However, neither of these pointings lie within the capped areas
of the chimney outflows and the O{\sc vi} is observed to be at lower
velocities than that of the shell.  Although it is likely that the
shell is filled with hot gas, no connections between the FUSE O{\sc
  vi} detections and the morphology of GSH 242-03+37 can be made at
this time.
\section{Discussion}
\label{sec:discussion} 
GSH 242-03+37 is a remarkable structure.  Its size, age, and
morphology are all at the extreme range of observed Galactic shell
parameters.  The only known Galactic supershell to display similar
properties is GSH 277+00+36 \citep{mcgriff03b}.  The similarities
between these two shells are intriguing.  Both shells have radii of
350 - 500 pc and expansion energies on the order of $10^{53}$ ergs.
Both are located far from the Galactic center at Galactocentric
distances of $\sim 10$ kpc.  The morphology of the two shells,
particularly their walls and extended $z$ structure, are strikingly
similar \citep[c.f.][Figure 1]{mcgriff03b}.  Although GSH 277+00+36 is
a much less confusing structure than GSH 242-03+37, it exhibits
similar chimney break-outs above and below the plane, as well as the
same scalloped structure along the walls.  The similar morphology of
these two objects is intriguing and leads us to speculate that their
large scale features may be dominated by common global Galactic
phenomena, for example the \HI\ scale height of the Galactic disk
(R. Sutherland 2005, private communication), rather than local
phenomena, such as the distribution of powering stars.  It should be
possible to determine which effects dominate with high resolution MHD
simulations of supershell evolution.  GSH 242-03+37, because it is
closer, offers some advantages over GSH 277+00+36.  In GSH 242-03+37
we can observe very weak chimney caps and explore the stellar content.
Here we discuss some of the characteristics that are specific to GSH
242-03+37.

\subsection{Stellar content}
\label{sec:stars}
\begin{figure}
\centering
\plotone{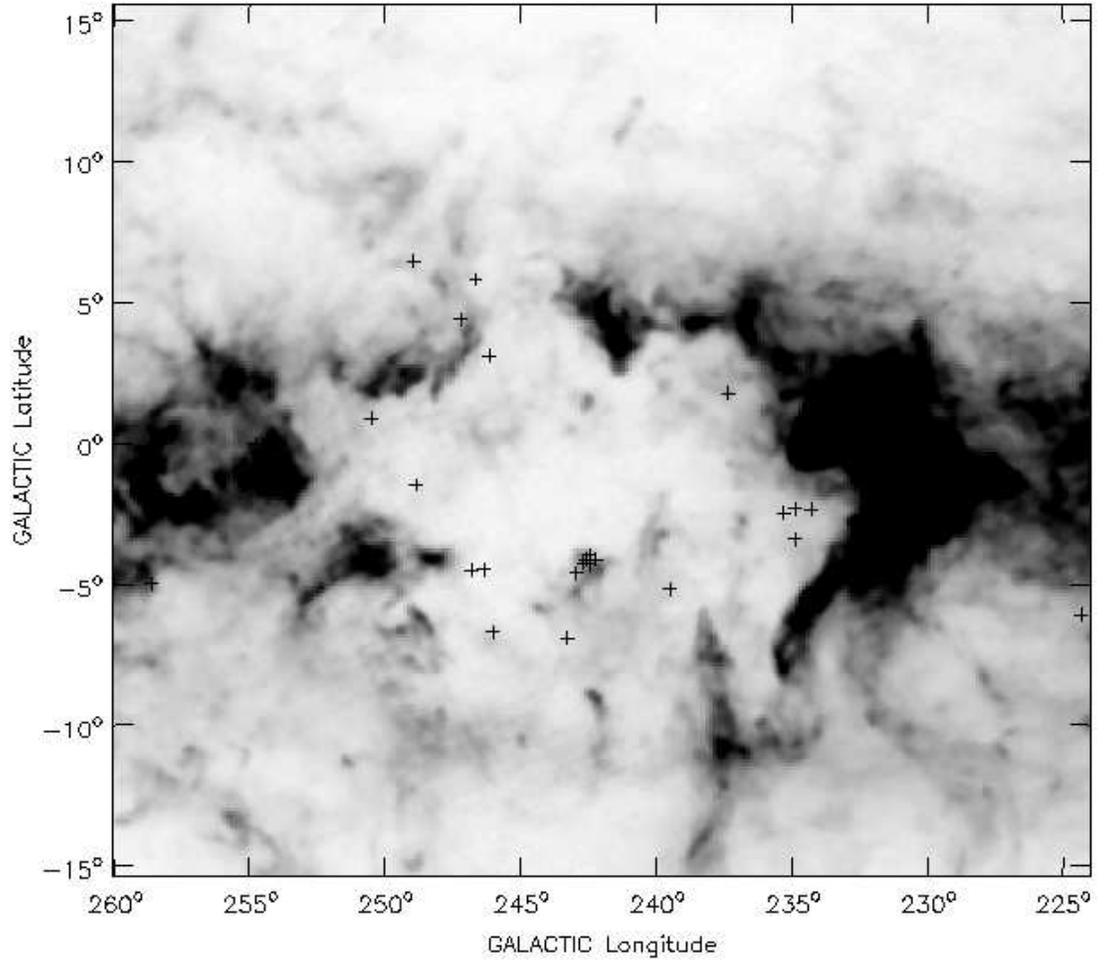}
\caption[]{\HI\ velocity channel image of GSH 242-03+37 at $v=45.3$
  \kms.  The grey scale in this image has a scaling power of $-0.4$
  and runs between 0 K (white) and 40 K (black).  Stars from
  \citet{kaltcheva00} located in the volume of the shell are plotted
  as crosses.
\label{fig:stars}}
\end{figure}

\begin{figure}
\centering
\plotone{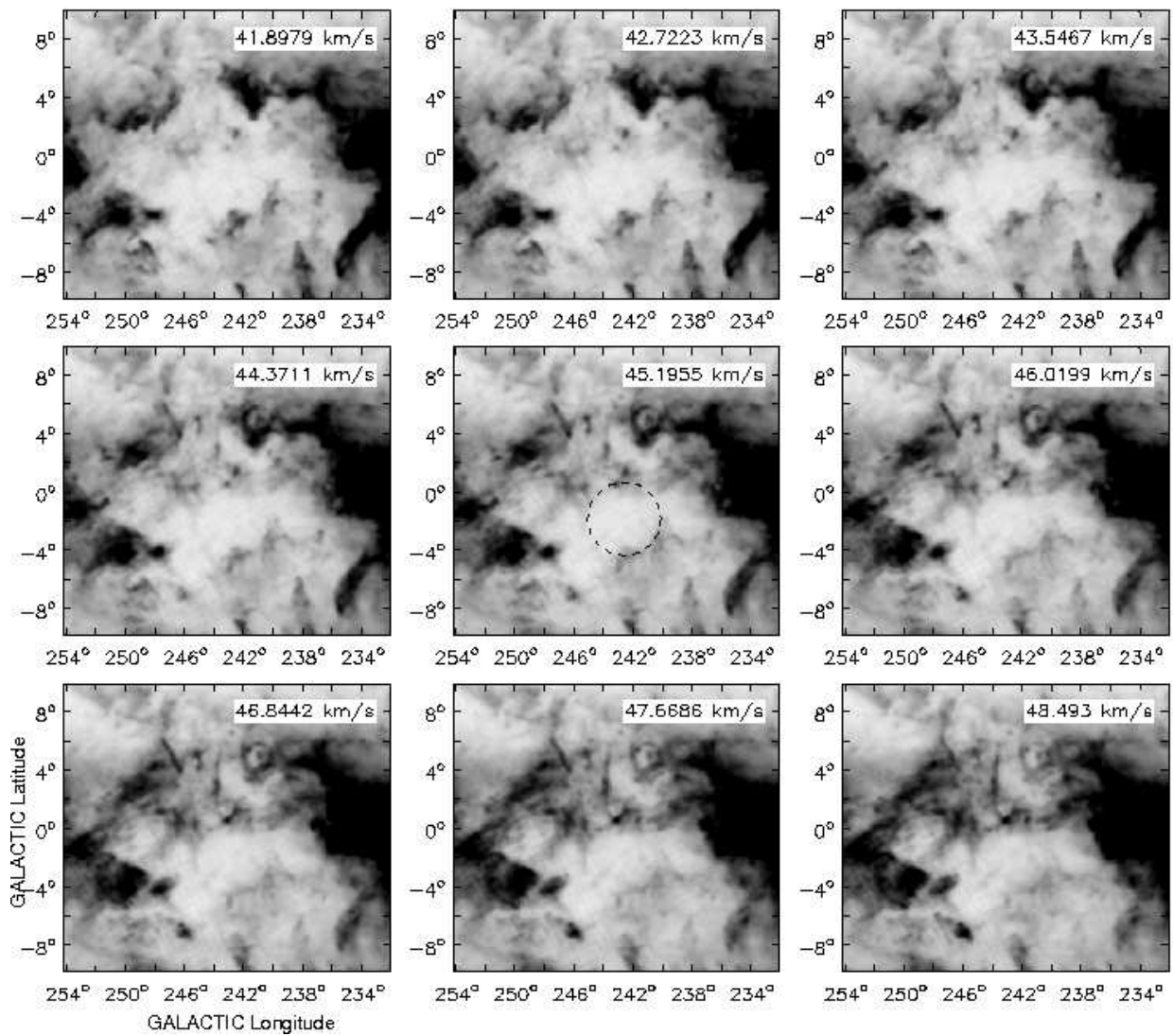}
\caption[]{The central region of GSH 242-03+37.  The grey scale runs
  from 0 K to 40 K with a scaling power cycle of -1.  The 
  image shows the ``mini-shell'', as a thin ring surrounding a void of
  $3\fdg8$ diameter centered on $l=242\fdg9$, $b=-2\fdg3$.  The
  mini-shell is outlined with a dotted line in the center panel.  The shell
  does not change size with velocity, but is apparent across at least
  6.5 \kms\ of velocity width.
\label{fig:minishell}}
\end{figure}

\begin{figure}
\centering
\includegraphics[width=5in]{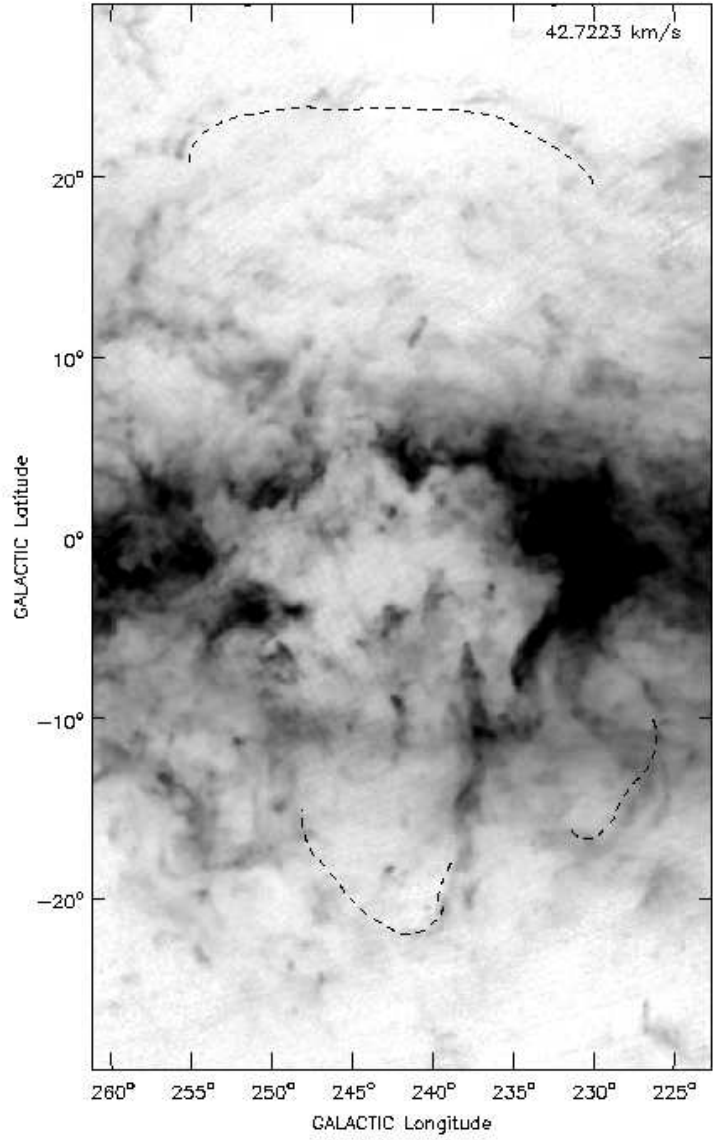}
\caption[]{\HI\ velocity channel image of GSH 242-03+37 at $v=45.3$
  \kms.  The grey scale in this image is logarithmic and scaled
  between 0 K (white) and 45 K (black).  The caps of the chimneys are
  marked with dotted lines.  The most obvious cap is visible above the
  shell. All caps are located $\sim 1.6$  kpc from the center of the shell.
\label{fig:annotate}}
\end{figure}

The stellar content of Galactic \HI\ supershells is largely unknown.
Most shells lie in the Galactic plane behind several magnitudes of
visual extinction, precluding most optical stellar surveys.  Infrared
surveys, such as 2MASS, can probe to much greater distances than in
the optical but the line-of-sight confusion of stars makes it
difficult to associate specific stars with supershells.  As it is, most of the
known OB associations in the Galactic plane are at distances of less
than 3 kpc.  By contrast, most \HI\ shells are at distances larger
than 2 kpc because the confusion of gas at local velocities
makes detecting shells difficult.  Fortunately GSH 242-03+37 is
located in the ``Puppis Window'' and benefits from numerous stellar
studies.  \citet{kaltcheva00} recently compiled and updated the
$uvby\beta$ photometry of luminous OB stars in the Puppis-Vela region,
providing a nearly complete table of the stellar types and distances.
We have extracted from their table all stars with projected distances within 1
kpc of the center of the shell. For the region $234 \leq l \leq
252\arcdeg$, $-7 \leq b \leq +7\arcdeg$ there are 22 OB stars, of
which the earliest is an O9 type star.  These stars are plotted as
crosses on Figure \ref{fig:stars}.

As one might expect for an old shell, there are very few massive stars
in the center of the object, with the notable exception of a small
cluster near $(l,b)=(242\arcdeg,-5\arcdeg)$.  These stars belong to a
cluster of O and B-type stars, which appear to lie along the rim of
the internal mini-shell.  The remaining stars are located near the
shell walls.  Along the right wall the stars appear to lie near the
foci of the loops that characterize the scalloped structure of GSH
242-03+37.  This coincidence seems to suggest that the stars near the
edge of the shell are contributing to the continued expansion of the
shell and determining the morphology of the walls.  In addition, the
stars trace the left-hand edge of the upper chimney wall, extending to
$b=6\fdg5$ or $z\approx 400$ pc at a distance of 3.6 kpc.  The mean
height for Galactic OB stars is only $90$ pc \citep{miller79}, so OB
stars at a height of 400 pc are unusual.  One obvious explanation for
their position is that they were formed out of dense
material raised to high $z$ by the shell.

A potential tracer of the past population of massive stars is the
current population of pulsars.  \citet{perna04} examined the number of
pulsars within several of the largest supershells in the Milky Way,
including GSH 242-03+37.  They compared the numbers of known pulsars
with Monte Carlo simulations of the pulsar population to predict how
many pulsars should be associated with a shell, assuming a multiple
supernovae formation scenario for the shell.  Although there are only
2 known pulsars within GSH 242-03+37, based on the current sensitivity
of pulsar searches they predicted that there should be 7 pulsars
within the shell.  The known pulsars therefore provide very few
constraints on the progenitor stars that may have formed the
supershell.

\subsection{The Mini-shell}
\label{sec:minivoid}
Figure \ref{fig:minishell} shows the central region of GSH 242-03+37.
The mini-shell (described in \S \ref{sec:morphology}) is apparent at
at $(l,b,v)$ = $(242\fdg9,2\fdg3,45~{\rm km~s^{-1}})$.  If it, too,
is at a distance of 3.6 kpc then the mini-shell has a radius of
$R_{\rm sh} \approx 120$ pc.  There is no clear indication that the
structure changes size with velocity although it is persistent over at
least 7 \kms\ of velocity width. The shell is likely a stationary
structure or one with a very small expansion velocity.  From our data
we can only estimate an upper limit to the expansion velocity of
$v_{\rm exp}\leq \Delta v /2 \sim 3$ \kms, which is less than the
turbulent velocity of warm \HI\ clouds, typically $\sim 7$
\kms\ \citep{belfort84}.  It is therefore impossible to distinguish
the shell's expansion from random cloud motions in the ISM.  For a
stationary shell a rough estimate of the age of the shell can be made
from the sound crossing time for a 120 pc radius.  Assuming $c_s \sim
10$ \kms\ for $T\sim 8000$ K \HI, the age is $t \sim R_{\rm sh}/c_s = 12$
Myr.

How does the mini-shell relate to GSH 242-03+37?  From the
\HI\ velocity channel images it appears as if the mini-shell is
located within GSH 242-03+37.  It is difficult to understand how a
cool neutral shell could form within a mostly evacuated supershell.
We have compared our estimates of the internal density of the
supershell with the swept-up mass of the mini-shell showing that there
is not enough gas in the supershell to form the mini-shell.  From the
measured column density along the mini-shell walls we estimate that
the typical \HI\ density in the walls is only $n_H \sim 0.4~{\rm
  cm^{-3}}$, a factor of a few lower than typical ISM values
\citep{dickey90}.  This gives a swept-up mass for the shell of $\sim 7
\times 10^4~{\rm M_{\odot}}$.  If the shell were formed near the
center of the main void, where we estimated that the typical
\HI\ density is only $n_H \sim 0.07~{\rm cm^{-3}}$, then the total
amount of mass enclosed in a sphere of radius 120 pc is only $\sim 1.2
\times 10^4~{\rm M_{\odot}}$, a factor of six less than the swept-up
mass of the shell.  If the excess mass in the mini-shell came from
swept-up ionized gas that had since recombined and cooled it would
imply an ionized density of $\sim 0.3~{\rm cm^{-3}}$.  This is much
higher than the typical ionized densities in shell interiors, which
are usually on the order of $\sim 5 \times 10^{-3}~{\rm cm^{-3}}$
\citep{maclow88}.  An alternative explanation is that the mini-shell
is a very old structure that was formed through the stellar winds of
the stars whose supernovae shocks eventually contributed to the large
shell.  In this case, the cool walls of the mini-shell might have been
overtaken by the supernovae shocks, which subsequently expanded to
much larger radii.  The size of the mini-shell is approximately
consistent with a late stage stellar wind bubble.  A final suggestion
is that the mini-shell formed not near the center of the shell, as it
seems in the projected image, but at the edge or outside of the
supershell where the gas densities should be higher.  In order to be
fully outside the main shell the mini-shell would require a systematic
velocity that is $\sim 5 - 10$ \kms\ different from the LSR motion at
its position.  Given the dynamical nature of a large shell, that kind
of motion is not unreasonable.  Although it seems unlikely that the
mini-shell formed in the interior of a swept-out supershell it is not
possible with our data to distinguish between the latter two
scenarios.

Along the lower wall of the mini-shell lies a small cluster of six O
\& B type stars \citep*{kaltcheva00,kaltcheva01}.  The cluster is at a
distance of $3.2\pm0.2$ kpc and contains the stars: LS 538 (B0II), 514
(B1II) 507 (B0.5 III), 534 (B3), 528 (O9III), and 511 (B2)
\citep{kaltcheva00}.  These stars are coincident with the brightest
part of the mini-shell.  Of these stars, only the later-type B2 and B3
stars are still on the main-sequence.  If we assume coeval star
formation, the age of the cluster must be between $\sim 7$ and 15 Myr
\citep{schaller92}, which is comparable to the age estimate for the
mini-shell.  There are two possible scenarios to explain the presence
of a stellar cluster on the edge of this internal shell.  The first is
that the mini-shell formed from the stellar winds or supernovae of a
single very massive star that was part of the stellar cluster.  This
seems unlikely if the stars are coeval, as the age of the stellar
cluster is comparable to or less than the age of the
shell. Alternately, the cluster may have formed out of gas compressed
on the edge of the expanding mini-shell, causing a new generation of
stars along the walls of the mini-shell.  For as well as we can
determine the ages of the shell and the cluster, this scenario seems most
likely.

Many studies have searched for evidence of triggered star formation in
supershells.  In a recent study of the W3/W4 complex \citet{oey05}
found evidence for three generations of star formation.  They point
out that, statistically, a hierarchical system of three or more
generations of star formation is more suggestive of a causal
relationship between the generations than a two generation system.  In
the GSH 242-03+37 system there is evidence of multiple epochs of star
formation, contributing respectively to the $\sim 21$ Myr old
supershell, the OB stars near the edges of the supershell, the
mini-shell and the stellar cluster along the edge of the mini-shell.
Whether these multiple epochs of star formation represent multiple
generations or simply continuing star formation is not clear.
Unfortunately, the ambiguous position of the mini-shell and the
uncertainties in the shell ages makes it extremely difficult to 
identify three unique generations of star formation in the system.  We
therefore state that the system is {\em suggestive} of triggered star
formation, but that the evidence is not conclusive.

\subsection{Chimneys and Halo Clumps}
\label{sec:chimneys}
GSH 242-03+37 has three chimney break-outs with the dominant one
towards positive latitudes.  All three chimneys are capped
approximately 1.6 kpc above the center of the shell.  The morphology
of the positive latitude chimney in particular is reminiscent of the
models of chimney formation, such as
\citet*{maclow89,tomisaka86}. \citet{maclow89} showed that for a
superbubble expanding in a Galactic disk with an exponential
atmosphere the shell will extend far beyond the Galactic mid-plane,
developing a polar cap at $z\sim 1500$ pc.  In the late stages of
shell evolution gravitational acceleration dominates the dynamics of
the slowly expanding shell and the polar cap should become
Rayleigh-Taylor unstable and fragment.  It is this fragmentation that
allows hot gas filling the shell cavity to escape to the halo
\citep{dove00}.  As seen in Fig.~\ref{fig:annotate} the upper cap of
GSH 242-03+37 shows some evidence for fragmentation.  There are a
number of dense concentrations along the general arc of the cap, as
well as some regions where the arc appears absent.  We note however,
that even with the sensitivity of GASS, the brightest features along
this arc are only a few Kelvin in brightness temperature, so very faint
portions of the arc may not be detectable.

If, as these observations indicate, the polar caps of expanding
supershells can reach heights of $z \sim 1500$ pc before fragmenting
it raises questions about the ultimate fate of the fragmented caps and
the expected size distribution for forming clouds.  These questions
have been addressed in a general sense in a variety of numerical
simulations, \citep[e.g.][]{avillez00,avillez01}.  \citet{avillez00},
for example, predicts that expanding shells should produce
condensations of size 5 - 100 pc on timescales of tens of million
years.  The simulations explain the cloudlets in terms of expelled
chimney gas that has cooled and recombined.  GSH 242-03+37, on the
other hand, seems to be producing cool clouds from the fragmented
shell, a process that presumably takes place well before expelled
chimney gas can cool.

The cool shell of GSH 242-03+37 appears to be fragmenting into
cloudlets with sizes of a few tens of parsecs.  The linewidths of
these clumps are on the order of $\sim 10~{\rm km~s^{-1}}$, indicating
thermal temperatures of $T\sim 10^3$ K or lower.  We measure column
densities (assuming they are optically thin) for these clumps of $N_H
\sim$ few $\times 10^{19}~{\rm cm^{-2}}$.  If the clumps are roughly
spherical, then their average \HI\ number density is $n \sim 1~{\rm
  cm^{-3}}$ and their \HI\ mass is $\sim 100~{\rm M_{\odot}}$.  The
\HI\ mass of the clumps is well below the dynamical mass limit to be
gravitationally bound, which is $\sim 2 \times 10^5~{\rm M_{\odot}}$
for a 10 pc cloud with a 10 \kms\ linewidth.

Are these clumps in pressure equilibrium with their surroundings?  The
thermal pressure of these clumps is $nT \sim 10^3~{\rm cm^{-3}~K}$.
The thermal pressure of the lower halo at a $z$-height of 1.6 kpc is
uncertain.  By mass and volume the dominant component of the ISM at
$z\sim 1.6$ kpc is warm ionized gas \citep{ferriere01}, with a density
of $\sim 4\times 10^{-3}~{\rm cm^{-3}}$ and a temperature of $\sim
8000$ K \citep{reynolds91}.  We also know from observations of the
soft X-ray background \citep{snowden98} and O{\sc vi} absorption
\citep{savage03}, among others, that there is a significant diffuse,
hot ($T\sim 10^{5-6}$ K) component to the lower halo gas.  This
component is difficult to observe directly but from FUSE O{\sc vi}
measurements \citet{savage03} suggest that it is distributed as a
patchy, plane-parallel exponential with a scale height of $\sim 2.3$
kpc.  The contribution of the hot, ionized medium to the thermal
pressure of the lower halo is a matter for debate with estimates
ranging from $\sim 10~{\rm K~cm^{-3}}$ \citep{boulares90} to $\sim
10^3~{\rm K~cm^{-3}}$ \citep{shull94}.  It is generally agreed,
however, that the presence of this medium, as well as its patchy
nature are probably due to exhausting hot gas from supershells like
GSH 242-03+37.  Therefore, the best estimate for the thermal pressure
of the ambient medium most likely comes from estimates of the thermal
pressure in the interior of an evolved supershell.

It is not trivial to estimate the thermal pressure in the interior of
GSH 242-03+37 because the shell has begun to break-out, releasing its
pressure and also because the shell is sufficiently evolved that
radiative cooling is important in the interior.  We can, however,
roughly estimate the maximum internal thermal pressure before
break-out, assuming an adiabatic interior where the internal density
is dominated by mass evaporated from the cold dense shell
\citep{maclow88}.  The internal thermal pressure for an evolved
spherical shell of age, $t_7 = t /10^7$ yr, formed with an energy
deposition rate of $L_{38} \approx E_E/t /(10^{38}~{\rm erg~s^{-1}})$
is $nT = (1.4 \times 10^4~{\rm K~
  cm^{-3}})\,L_{38}^{14/35}n_0^{21/35}t_7^{-28/35}$ \citep{maclow88}.
If, once again, we assume that the ambient density is $n_0 \sim 1~{\rm
  cm^{-3}}$, then the internal pressure is $\sim 1.4 \times 10^4~{\rm
  K~cm^{-3}}$.  Given that the shell has begun to break apart and that
its $z$-height far exceeds its dimension along the Galactic plane,
this pressure estimate is almost certainly too large but it provides a
useful limit to the pressure around the clumps.  We may therefore
conclude that the clumps are likely in equilibrium or moderately
pressure-supported by the hot gas from the shell.

We would like to know whether and for how long clouds created from the
fragmentation of a shell could survive against evaporation due to heat
flux from the ambient medium.  If the ambient medium is indeed
dominated by warm ionized gas, with $T\sim 8000$ K, then the classical
thermal evaporation is extremely long when compared to the 20 - 30 Myr
lifetime of the shell \citep{cowie77}.  Even if the ambient medium is
dominated by the hot, diffuse gas of the shell interior, the
evaporation time is $\sim 35$ Myr.  We can expand on this estimation
somewhat by following \citet{mckee77b} who consider the simple
scenario of cool, spherical clouds embedded in a hot medium where the
fate of the cloud is controlled by the effects of both radiative
losses and incoming heat flux. Those authors provide analytic
solutions to determine the critical radius, $R_{cr}$ below which
clouds evaporate and above which they condense material from the
surrounding medium.  The critical radius is determined solely by the
pressure in the ambient medium.  From Figure 2 in \citet{mckee77b} we
estimate that the critical radius for $T\sim 8000$ K gas with a
density of $\sim 4\times10^{-3}~{\rm cm^{-3}}$ is $\sim 3$ pc.  The
clouds, however, are also affected by the very hot, tenuous medium
interior to the shell.  Even if this gas is at $\sim 10^6$ K with
densities of $10^{-2}~{\rm cm^{-3}}$ the critical radius will be on
the order of 15 pc and even lower for smaller ambient densities.
Given these very crude assumptions it seems that the clouds should be
near or above the critical radius for all expected temperatures in
their environs.  These clouds should therefore be relatively
long-lived and may even condense matter onto themselves.

Another interesting question is what size scales should we expect to
see represented from fragmenting caps.  This of course depends on
the instability process responsible for the fragmentation.  If the
shell cap fragments through classical Rayleigh-Taylor instabilities
all size scales are expected to be represented, but the growth
timescale of the instability is proportional to the square-root of the
size scale so we should expect to see the smaller scales first.  In
addition, in the presence of a magnetic field a lower limit is applied
to the size of growing modes and also a fastest growing mode is
established \citep{mcgriff03b}.  In that case, the size scales
observed in supershells may provide probes of the ambient medium.

In GSH 242-03+37 we observe large polar caps that appear to be
breaking into clumps with radii on the order of tens of parsecs.  The
size, density, linewidths and $z$-height of these clumps is very
similar to halo cloudlets detected by \citet{lockman02a} and those
found in a recent study of the GASS pilot region (A.\ Ford et
al.\ 2005a, in prep.).  The origin of the \citet{lockman02a} clouds is
still quite uncertain.  If the polar caps of supershells like GSH
242-03+37 can break into small clumps with parsec or tens-of-parsec
size scales these clumps should be much longer lived than the shell
itself.  Although these ideas are still rather speculative, the
similarities of properties suggests that it would be worth pursuing
the fragmenting shell model further.  Some important questions to
answer will be: how long can the clouds survive? what is their $z$
distribution? and, given that they are massive compared to their
surroundings, how long before they will drop back to the Galactic
plane?  We will address these questions in a future paper comparing
the properties of halo cloudlets with simulations of the long term
evolution of supershells (A.\ Ford et al.\ 2005b, in prep.).

\section{Conclusions}
\label{sec:conclusions}
We have presented new \HI\ images of the Galactic supershell GSH
242-03+37 from the Galactic All-Sky Survey (GASS).  GSH 242-03+37 is
one of the largest shells in the Galaxy with a radius of $R_{sh} =
565\pm 65$ pc.  We show that the supershell is broken at the edge of
the disk, both above and below the plane.  The resultant structure has
three ``chimney'' openings that are capped with very narrow filaments
all situated $\sim 1.6$ kpc above the disk midplane.  These ``caps''
are extremely reminiscent of the caps seen on expanding supershells in
simulations, such as those by \citet{maclow89} and \citet{tomisaka98}.
In supershell evolutionary theories these shells should become
Rayleigh-Taylor unstable and the polar caps break into clumps.  The
caps of GSH 242-03+37 appear to show clump structures with sizes on
the order of 20 pc, which may indicate the onset of break-out.  We
estimate that clouds formed through this break-out may survive longer
than the parent shell.  The size, temperature and $z$-heights of these
clouds are similar to the halo cloudlets detected near the disk in the
inner Galaxy \citep{lockman02a}.  We suggest that the Lockman
cloudlets may be formed through the fragmentation of high-$z$
supershell caps.  All-sky surveys like GASS will provide an extremely
valuable database for testing this idea.  GASS will have the  sky
coverage, resolution and sensitivity necessary to detect and study the
relationship of small-scale structures in the halo to structures in the
Galactic disk.

We have searched catalogs of OB stars for massive stars in the
vicinity of GSH 242-03+37.  We find very few stars at the center of
the shell, but there are 22 OB stars that lie near the internal edges
of the shell, of which the earliest is an O9 type star. There are six
OB stars with ages between 7 and 13 Myr \citep{kaltcheva01} that lie
along a small ``mini-shell'' that looks as if it is inside GSH
242-03+37.  It is difficult to understand how a neutral shell could
form in the evacuated cavity of GSH 242-03+37.  We therefore suggest
that it lies at the edge of the shell and that the OB stars were
formed in material compressed along the walls of the mini-shell.  The
agreement between the main shell structure, the identification of 22
OB stars near the shell walls, the mini-shell and its corresponding
cluster of young OB stars are suggestive, but not conclusive evidence
for triggered star formation.

\acknowledgements The Parkes Radio Telescope is part of the Australia
Telescope which is funded by the Commonwealth of Australia for
operation as a National Facility managed by CSIRO.  This research was
performed while D.J.P.\ held a National Research Council Research
Associateship Award at the Naval Research Laboratory. Basic research
in astronomy at the Naval Research Laboratory is funded by the Office
of Naval Research.  D.J.P. also acknowledges generous support from NSF
MPS Distinguished International Research Fellowship grant
AST0104439. B.K.G. acknowledges the financial support of the Australian
Research Council through its Discovery Project program.  We are
extremely grateful to Warwick Wilson, March Leach, Brett Preisig, Tim
Ruckley and John Reynolds for their efforts in enabling the GASS
correlator mode in time for our first observations.


\clearpage

\begin{deluxetable}{ll}
\tabletypesize{\footnotesize}
\tablecaption{Basic parameters for GSH 242-03+37
\label{tab:params}}
\tablehead{
\colhead{Parameter} & \colhead{Value}
}
\startdata
Center $l$ & $243\arcdeg$ \\
Center $b$ & $-1\fdg6$ \\
Central $v_{\rm LSR}$ & $+42$ \kms \\
Distance & $3.6\pm0.4$ kpc\\
Radius & $565\pm 65$ pc \\
Velocity FWHM & 25 \kms\\
$v_{\rm exp}$ & 7 \kms ($20~{\rm km~s^{-1}}$)\tablenotemark{a}\\
$E_{\rm E}$ & $3 \times 10^{53}$ ergs ($1.6 \times 10^{54}$ ergs)\\
\enddata
\tablenotetext{a}{Values given in brackets are from \citep{heiles79}
  and are only given where they differ significantly from the values
  determined here.}
\end{deluxetable}

\end{document}